\documentstyle[12pt,epsf]{article}
 
 \def \dfrac #1#2 {\displaystyle\frac{#1}{#2}}
\begin{document}
 {\bf \begin{center}
 On the Extraction of the Neutron Spin Structure Functions and
 the Gerasimov -- Drell -- Hearn
 Integral from $^3\vec {He}(\vec e, e')X$ data
 \end{center} }

 \vskip 5mm

 \begin{center} C.  Ciofi degli Atti and
 S.  Scopetta \footnote{Now at { \sl Institut f\"ur Kernphysik der 
 Universit\"at Mainz, Joh.-Joachim-Becher-Weg 45, D-55099 Mainz, Germany}}
\\[2mm]
{\em Department of Physics, University of Perugia, and
INFN, Sezione di Perugia,
\\
via A. Pascoli, Perugia, I-06100, Italy.}
\end{center}

 \vskip 5mm

\baselineskip 21pt

\begin{abstract}
Nuclear effects in polarized inelastic electron scattering
off polarized $^3He$ are discussed;
in the kinematics of future experiments
at CEBAF, 
Fermi motion effects are found to be much larger than
in deep inelastic scattering. 
It is shown that improperly describing nuclear dynamics
would lead to the extraction of unreliable
neutron spin structure functions. On the other hand side,
a simple and workable
equation relating the Gerasimov -- Drell -- Hearn Integral 
for the neutron
to the corresponding quantity for $^3He$ is proposed.
\end{abstract}

\leftline{PACS: 13.60.-r, 25.10.+s, 11.55.Hx, 14.20.Gk}
\leftline{KEYWORDS: GDH integral, Few nucleon systems,
sum rules, nucleon resonances} 

\newpage
 
The measurement of the polarized nucleon Spin Structure Functions 
(SSF) $g_1$ and $g_2$ in the 
resonance 
region allows one to check the helicity structure 
of the photon -- nucleon
coupling between the Deep Inelastic Scattering (DIS) region 
and the real photon limit~\cite{drech}.
Recently it has been proposed at CEBAF to
study the SSF in a wide range of 
energy ($0.2~GeV \leq \nu \leq 3~GeV$) and momentum
($0.15~GeV^2\leq Q^2 \leq 2~GeV^2$) transfers,
for both the proton \cite{cebp} and the neutron,
using in the latter case
polarized deuteron \cite{cebd} and $^3He$ \cite{ceb3}
targets.
   
For any spin ${1 \over 2}$ hadronic target $A$, the SSF $g_1^A$ 
reads as follows \cite{IOFFE}
\begin{equation}\label{g1}
g_1^A(\nu,Q^2)=\frac {M_A K}{8\pi^2 \alpha(1+\frac {Q^2}{\nu^2})} \left [
\sigma_{1/2}(\nu,Q^2)-\sigma_{3/2}(\nu,Q^2)+\frac {2\sqrt {Q^2}}{\nu}
\sigma_{TL}(\nu,Q^2)\right ]
\end{equation}
where $\sigma_{1/2(3/2)}(\nu,Q^2)$ is the cross section
for photon -- hadron scattering with 
total helicity ${1/2\,(3/2)}$,
$\sigma_{TL}(\nu,Q^2)$ is the interference between
the transverse and the longitudinal cross sections, $K$ is the
photon flux and $M_A$ is the hadron mass.

A relevant quantity related
to the SSF $g_1^N$ of the nucleon is the Gerasimov -- Drell -- Hearn (GDH)
In\-te\-gral: 
\begin{equation}\label{gdh}
I_{GDH}(Q^2)=
\int\limits^{\infty}_{\nu_{th}}
\frac {d\nu}{\nu}
\left ( {\sigma_{1/2}(\nu,Q^2)-\sigma_{3/2}(\nu,Q^2)}
\right ),
\end{equation}
where $\nu_{th}=(Q^2+2m_{\pi}M+m_{\pi}^2)/2M$ is the threshold 
energy for the
pion-e\-le\-ctro\-pro\-du\-ction, $m_{\pi}$ is the pion mass and
$M$ is the nucleon mass.
The integral $I_{GDH}$ gives in the real photon limit the GDH
Sum Rule \cite{DHG66}:
\begin{eqnarray}\label{sr}
I_{GDH}(Q^2=0) & = &
\int\limits^{\infty}_{\nu_{th}}
\frac {d\nu}{\nu}
\left ( \sigma_{1/2}(\nu,Q^2=0)-\sigma_{3/2}(\nu,Q^2=0) \right ) 
= -\frac {2 \pi ^2 \alpha }{M^2}\kappa^2
\cr 
& \simeq & 
\cases{
-0.53 \quad GeV^{-2} & 
for protons\cr
-0.60 \quad GeV^{-2} & for neutrons \cr}
\end{eqnarray}
where $\kappa$ is the anomalous magnetic moment of the nucleon.

Using (\ref{g1}) and (\ref{gdh}),
$I_{GDH}(Q^2)$ can be related in the large $Q^2$ limit
to the first moment of the SSF $g_1$ and,
by using the EMC result for the DIS SSF of the
proton $g_1$ \cite{EMC},  one gets, around
$Q^2 \simeq 10 ~GeV^2$,
$I_{GDH}(Q^2)
\simeq 0.14/Q^2$,
which is evidence of large changes
in the helicity structure of the $\gamma p$ coupling between the real
photon limit and the DIS region, leading to a change of sign of
$I_{GDH}(Q^2)$ at some value of $Q^2$.
Much theoretical work has been produced,
both in the real photon limit \cite{ik73} and at  
finite values of
$Q^2$ \cite{burk}, in order to understand this behavior,
which should be mainly due to the electroexcitation 
of nucleon resonances.

Extensive experimental 
programs are planned
to investigate not only the
$Q^2$ evolution of the
GDH Integral, but also, for the first time,
the proton and neutron 
GDH Sum Rule \cite{ar};
in the latter case,
several analyses of the available 
unpolarized photoproduction data do not agree 
with the prediction (\ref{sr}) for the neutron,
while similar estimates seem to give the correct GDH value
for the proton; it is for this
reason that the neutron
measurement is of particular relevance \cite{drech}. 

It is worth noticing that the measurement of the neutron SSF
in the CEBAF experiments will also 
contribute to the investigation of the 
low $Q^2$ evolution of the
Bjorken sum rule, which requires an accurate knowledge
of $g_1^n$ for $0 \leq x \leq 1$, being $x = Q^2 / 2M\nu$
the Bjorken variable.

Nuclear effects on the SSF
$g_1$ of the deuteron have already been studied 
\cite{plb} in the kinematics of the planned CEBAF
experiment; the aim of this
letter is to present the results of calculations
of nuclear effects on the spin structure function of $^3He$.

\indent 
A convolution model for the
nuclear SSF $g_{1(2)}^A$ for any spin ${1 \over 2}$ nucleus
has been obtained in
\cite{pg,sa}; 
it reads as follows
\begin{eqnarray}
 g^A_1   \left( Q^2,\nu\right ) & = & 
 \sum\nolimits\limits_{N=p,n} \int \nolimits \nolimits~dz
\int \nolimits \nolimits~dE \int \nolimits
\nolimits~d{\vec p}~ {M \over E_p}~{M \nu  \over p \cdot q}~
g_{1}^{N}\!\left(z,\nu,{Q}^{2} \right) 
\left \{
~{\hbox{\Large\it P}}_{\parallel}^{~N}\!\left( {\vec p},E\right)
\right. 
\cr
& & 
\left. 
+~{\hbox{\Large\it T}}^{~N}\!\left( {\vec p},E,Q^2\right) \right \} 
\delta \left(z+{M^2-p \cdot p\over 2M \nu} -{q \cdot p \over
M \nu}\right) 
 \label{eq11}
\end{eqnarray}

\noindent where $E$ is the removal energy,
$p\equiv(M_A-\sqrt{(E+M_{A}-M)^2+|\vec
p|^2},~\vec p)$
is the 4 -- momentum of the bound nucleon, $~E_p=\sqrt{M^2+|\vec{p}|^2}$
and $g_{1}^{N}\!\left
(z,\nu, Q^{2} \right)$ is the nucleon SSF.
Nuclear effects are described by the quantities 
${\hbox{\Large\it P}}_{\parallel}^{~N}\!\left(  \vec {p}, E\right)$ 
and ${\hbox{\Large\it T}}^{~N}\!\left(\vec {p},E, Q^2\right)$,
which are both related to the elements of the 2x2 matrix,
representing the spin dependent spectral function of a nucleon inside a
nucleus with  polarization $\vec{S}_A$ ~\cite{prc}. 
The elements of this matrix are
\begin{eqnarray}
P_{\sigma,
\sigma',\cal{M}}^{N} ({\vec{p},E})=\sum\nolimits\limits_{{f}_{A-1}}
{}~\langle{\vec{p},\sigma;\psi
}_{A-1}^{f} |{\psi }_{J\cal{M}}\rangle~ \langle{\psi
}_{J\cal{M}}|{\psi }_{A-1}^{f};\vec{p},\sigma '\rangle ~
\delta (E-{E}_{A-1}^{f}+{E}_{A})
\label{eq9}
\medskip
\end{eqnarray}
where $|{\psi}_{J\cal{M}}\rangle$ is the ground state of the target 
nucleus polarized along
$\vec{S}_A$, $|{\psi }_{A-1}^{f}\rangle$ is an eigenstate of the (A-1) nucleon
system corresponding to an
energy $E_{A-1}^f$, $|\vec{p},\sigma\rangle$ 
is the plane wave for the nucleon $N\equiv p(n)$.
In particular, ${\hbox{\Large\it P}}_{\parallel}^{~N}\!\left(\vec {p} ,
E\right) = 
P_{ {1 \over 2} {1 \over 2}, {1 \over 2} }( \vec p, E)
- P_{ - {1 \over 2} - {1 \over 2}, {1 \over 2} }( \vec p, E)$,
whereas the $Q^2$ dependent term 
${\hbox{\Large\it T}}^{~N}\!\left(\vec {p} ,
E, Q^2\right)$,
which gets contribution from both $g_1^N$ and $g_2^N$,
depends also upon 
$P_{ {1 \over 2} -{1 \over 2}, {1 \over 2} }( \vec p, E)$;
this term has been found
to give a very small
contribution in both the Deep Inelastic Scattering 
\cite{pg} as well as in the present
calculation of the resonance region:
for this reason it will be omitted hereafter.
In the DIS kinematics
($Q^2 \rightarrow \infty$, $\nu \rightarrow \infty$), Eq. (\ref{eq11})
gives the simple convolution formula described in Ref \cite{pg}:
\begin{eqnarray}
g_1^A(x) &  = &  \sum_N \int \limits_x ^A dz
~{1 \over z}~ g_1^N \left( {x \over z} \right)
~G^N(z)~~, \label{fin}
\medskip
\end{eqnarray}
with the light -- cone momentum distribution given by
\begin{eqnarray}
G^N(z) & = &  \int dE\, \int d {\vec p}~
{\hbox{\Large\it P}}_{\parallel}^{~N}\!\left({\vec p},E \right) 
\delta \left(z - {p^+ \over M} \right)~~~ \label{lux}
\medskip
\end{eqnarray}
where $p^+=p^0- \vec p \cdot \vec q / |\vec q|$ is the light -- cone
momentum component.

It is well known \cite{trial} that the impulse approximation used
above
is at variance by about 4 \% with
the experimental results on $\beta$ decay,
which are related to the integral of the difference of the SSF of $^3H$,
$g_1^{^3H}$, and $^3He$, 
$g_1^{^3He}$,
through the Bjorken Sum Rule for the three nucleon systems.
Recently \cite{fs} it
has been argued that nuclear shadowing at $x \leq 0.05$ 
may be responsible for such a disagreement, and could influence
$g_1^{^3He}$ up to $x\simeq 0.15$
This effect should not affect our results, which involve
larger $x$ values in the kinematics
of the planned experiments, at least for $Q^2 \geq 0.3 ~GeV^2$.
 
In the present letter the $^3He$ structure function $g_1^{^3He}$ 
has been calculated 
in the resonance region 
($W_{th}=M+m_{\pi}<\tilde W<2 ~GeV$,
being $\tilde W=\sqrt{(p+q)^2}$ the invariant mass of the virtual 
photon -- bound nucleon system),
by evaluating the following equation
\begin{eqnarray}
g^A_1   \left( x, Q^2\right ) & = & 
\sum\nolimits\limits_{N=p,n} ~ \int \limits_{W_{th}}
~d {\tilde W} ~{ \tilde W}
\int \nolimits \nolimits~dE \int \nolimits
\nolimits~d{\vec p}~ {M \over E_p E_X}~{M \nu  \over p \cdot q}~
g_{1}^{N}\!\left(x,{Q}^{2},\tilde W \right) 
\cr 
& &
{\hbox{\Large\it P}}_{\parallel}^{~N}\!\left( {\vec p},E\right)
\delta \left(\nu + M_3 -
E_R - E_X \right)
 \label{eqpa}
\end{eqnarray}

obtained from Eq. (\ref{eq11}) by changing the integration variable 
$z$ into $\tilde W=\sqrt{(p+q)^2}=\sqrt{2 M \nu z+M^2-Q^2}$,
being $E_X=\sqrt{ \tilde W ^2 + (\vec p + \vec q)^2}$ the energy
of the hadronic state produced by the interaction
of the struck nucleon with the incoming virtual photon, and
$E_R=\sqrt{(E+M_{A}-M)^2+|\vec
p|^2}$ that of the recoiling 2 -- body system.

The model of \cite{burk}
for the transverse virtual photon 
absorption cross sections $\sigma_{1/2}$ and 
$\sigma_{3/2}$ has been used to obtain the nucleon
SSF $g_1^N$ according to Eqs. (\ref{g1}).
In this model,
the contribution of the resonance $R$ is written
in the following way:
\begin{equation}
\sigma^R_{1/2(3/2)}(\nu,Q^2)=
{4M \over W_0 \Gamma_0}{A^{R~2}_{1/2(3/2)}}
(Q^2)\,B(\nu,Q^2)
\end{equation}
where $M$ and $W_0$ are the nucleon and resonance masses, 
$\Gamma_0$
is the total width of the resonance, $B(\nu,Q^2)$ 
is the extension to electroproduction of the
Breit -- Wigner 
parameterization given in
\cite{wal} for photoproduction,
$A^R_{1/2(3/2)}(Q^2)$ 
is the helicity amplitude for the excitations of the
resonances. In this model,
the amplitudes pertaining to the 
resonant states $P_{33}(1232)$, $D_{13}(1520)$,
$S_{11}(1535)$ and $F_{15}(1680)$ have been parametrized by using 
the existing experimental data, while other 
states have been added by using the
predictions of the Single Quark Transition Model; a non resonant
background due to the single pion Born term has also been included in 
the calculations. In the kinematics of the planned experiments the longitudinal
asymmetry can be approximated by the relation:
$
A^{^3He}(x,Q^2)  \simeq$ $2x {g_1^{^3He}(x,Q^2)/ F_2^{^3He}(x,Q^2)}
$
where $F_2^{^3He}(x,Q^2)$ is the unpolarized structure function of $^3He$.

Calculations have been performed in the resonance region for both
$g_1^{^3He}$ and $F_2^{^3He}$, using for the former
eq. (\ref{eqpa}) and for the latter the convolution approach
described in \cite{cdayl}. The results are 
presented in Figs. 1 and 2, where they are compared with those
obtained in DIS.
In the latter case, as it can be seen from Fig. 1 (a),
the following 
equation has been shown \cite{pg} to be a reliable approximation of the 
convolution formula, at least for $x \leq~0.8$:
\begin{equation}
g_1^{^3He}(x,Q^2)  \approx  2p_p g_1^p (x,Q^2) + p_n 
g_1^n(x,Q^2)
\label{gmod2}
\medskip
\end{equation}
where $~p_{p(n)}~$ is the effective nucleon polarization, produced by
the $S'$ and $D$ waves in the ground state of $^3$He
($p_{n}$$=1-{2 \over 3}P(S')-{4 \over 3}P(D)$,
$p_p=-{1 \over 3}(P(D)-P(S')$, being $P(D)$ and $P(S')$ the 
$D$- and $S'$- wave probabilities,
and given by
\begin{eqnarray} 
p_{N}& = & \int dE~\int d\vec p ~{\hbox{\Large\it
P}}_{\parallel}^{~N}\!\left(\vec p,E\right)~.
\label{pol}
\medskip
\end{eqnarray}
It can be seen from Fig. 1 (b) that an approximation similar to
Eq. (\ref{gmod2})
does not hold in the
resonance region, which means 
that nuclear effects are relevant; we found 
that binding effects
are very small, so that  
it is the Fermi motion which is responsible for
the relevant broadening and damping of the peaks associated
to the excitation of
the most prominent resonant states in the nuclear medium. This fact 
is a well known result also in the unpolarized case, as it can be seen in
Fig. 2.  Fig. 2 (a) shows that, in DIS kinematics, 
the unpolarized structure function $F_2^{^3He}$,
calculated by taking into account Fermi motion and binding,
hardly differs from the sum of the structure functions 
of the free nucleons,
whereas Fig. 2 (b) shows that 
in the resonance region ($Q^2=1~GeV^2$)
the resonant peaks are
strongly damped by Fermi motion.
On the other hand side, we found
that 
the proton contribution to the nuclear $g_1$
is small, 
both in the DIS and resonance regions, so that
in both regions
$\vec{^3He}$ is a nice effective polarized neutron
target.
  
In DIS, the quantity
\begin{eqnarray} 
\tilde{g}_1^n(x,Q^2)={1 \over p_n} \left[ 
g_1^{^3He}(x,Q^2)-2p_p g_1^p(x,Q^2) \right]~,
\label{g1art} 
\medskip 
\end{eqnarray}
obtained by inverting Eq. (\ref{gmod2}) and calculated
using the convolution formula for 
$g_1^{^3He}(x,Q^2)$, differs from the free neutron SSF $g_1^n$,
by at most $4\%$.
Such a small difference is rather
independent of the form of any well behaved $g_1^N$, 
and therefore Eq. (\ref{gmod2}) can be considered a workable formula for 
extracting $g_1^n(x,Q^2)$ from the experimental
$g_1^{^3He}(x,Q^2)$. 
The same doesn't hold in the resonance region, so that an
alternative way for such an extraction has to be figured out.
A possible solution is to use the
method described in \cite{kanna}, initially applied
in the deep inelastic region and extended in Ref. \cite{plb}
to the extraction of the neutron spin dependent structure functions
at finite $Q^2$ from the deuteron data.

We would like to stress that in this process
the $P_{33}(1232)$ ($\Delta$) resonance, which gives a relevant
contribution to the nuclear $SSF$, is produced with a large
average 3 -- momentum $|{\vec p}_\Delta|$. For example,
at $Q^2\simeq 1~GeV^2$, $|{\vec p}_\Delta| \simeq 6 ~f^{-1}$
and it remains large even
at $Q^2\simeq 0.5~GeV^2$ ($|{\vec p}_\Delta| \simeq 4.5 ~f^{-1}$).
Moreover, at any value of $Q^2$, heavier resonances are
excited with a 3 -- momentum larger than  $|{\vec p}_\Delta|$.
It means that the produced resonant states come out
from the nucleus carrying high kinetic energy with respect
to the two -- body spectator system; thus the final state
interactions, which have been disregarded in the present
impulse approximation approach, should not play
a relevant role.

Finally,
let us discuss the role of nuclear effects on the 
GDH Integral.
In the DIS limit, it has been observed \cite{pg}
that the integrals of the formula (\ref{eq11}) and that of
the approximation Eq. (\ref{gmod2}) differ by a negligible amount.
Then the integral $\Gamma^n$ of $g_1^n$ ($\Gamma^n = \int g_1^n(x)~dx$)  
can be easily obtained 
from the experimentally known $\Gamma^{^3He}$ and $\Gamma^p$ integrals,
by integrating Eq. (\ref{g1art}), obtaining
\begin{eqnarray} 
\label{itiln3}
\tilde{\Gamma}^n(Q^2)={1 \over p_n} \left[ \Gamma^{^3{\rm He}}(Q^2)
 - 2p_p {\Gamma}^p(Q^2) \right]~.
\end{eqnarray}
Let us check whether such a procedure can be applied to the resonance region.
To this end let us introduce, for any spin ${ 1 \over 2}$ target $A$,
the following integral

\begin{equation} 
\label{intnuc}
I^A(Q^2)={8 \pi^2 \alpha \over M}
\int\limits_{\nu_{th}}^\infty
{ d \nu \over \nu} 
{ \left ( 1 + {Q^2 \over \nu^2} \right ) \over K} 
g_1^A(\nu,Q^2)~,
\end{equation}
which, in the case of the nucleon, coincides with the GDH integral (\ref{gdh}),
provided the interference contribution is disregarded in 
the definition of $g_1^A(x,Q^2)$ (Eq. (\ref{g1})),
which will be assumed hereafter.
In Figure 3 we show the quantity 
\begin{eqnarray} 
\label{itiln3}
\tilde{I}^n(Q^2)={1 \over p_n} \left[ I^{^3{\rm He}}(Q^2)
- 2p_p I^p(Q^2) \right]~,
\medskip 
\end{eqnarray}
calculated using in (\ref{intnuc})
the model described in \cite{burk} 
to evaluate $g_1^p$ and the convolution
formula (\ref{eqpa}) to obtain
$g_1^{^3{\rm He}}$.
It can be seen that this curve differs at most by
$5 \%$ from the free neutron $I^n(Q^2)$,
obtained using in (\ref{intnuc}) 
the model given in \cite{burk} for $g_1^n$.
It appears therefore that
the simple formula (\ref{itiln3}) could be
used to get the neutron GDH Integral
from the measured $I^{^3He}(Q^2)$ integral.
This quantity is also shown in the Figure:
a comparison with the integral corresponding to the free
neutron, shows that nuclear structure effects are large,
but can be safely taken care of by simply
using the effective polarizations.

To sum up, our calculations show that in the resonance region
$g_1^{^3He}(x,Q^2) \neq p_ng_1^n(x,Q^2)$ $+2p_p
g_1^p(x,Q^2)$, but the integrals of these two quantities
are very similar.
Calculations with different types of $g_1^N$ would be very useful
in clarifying the model dependence of such a result.

We would like to express our gratitude to V. Burkert, L. P. Kaptari and
A. Yu. Umnikov for enlightining discussions.

\bigskip

\newpage
\centerline{\bf \large Captions}
\hfill\break
\hfill\break
\hfill\break
{\bf Figure 1}: $g_1^{^3He}$ 
in DIS ($Q^2=\,10\,GeV^2$) (a)
\protect \cite{pg}, and in resonance ($Q^2=1~GeV^2$) (b) regions,
obtained by considering Fermi
motion and binding (full). The dashed curve represents the same functions
obtained considering the proton and neutron effective polarization in $^3He$
as the only relevant nuclear effects (Eq. (\protect \ref{gmod2})).
\hfill\break
\hfill\break
{\bf Figure 2}: $F^{^3He}_2$ in DIS ($Q^2=\,10\,GeV^2$) (a) 
and in resonance ($Q^2=\,1\,GeV^2)$ (b)
regions, calculated by properly considering 
Fermi motion and binding (full curves) and by summing the unpolarized
structure functions of the nucleons (dotted curves).
\hfill\break
\hfill\break
{\bf Figure 3}: The integral ${\tilde I}^n(Q^2)$,
Eq. (\protect \ref{itiln3}) (crosses),
compared with $I^n(Q^2)$ \protect\cite{burk}
(full) and with 
$I^{^3{\rm He}}(Q^2)$, Eq. (\protect \ref{intnuc}) (dots).

\newpage

\begin{figure}[h]
\vspace{12cm}
\includegraphics{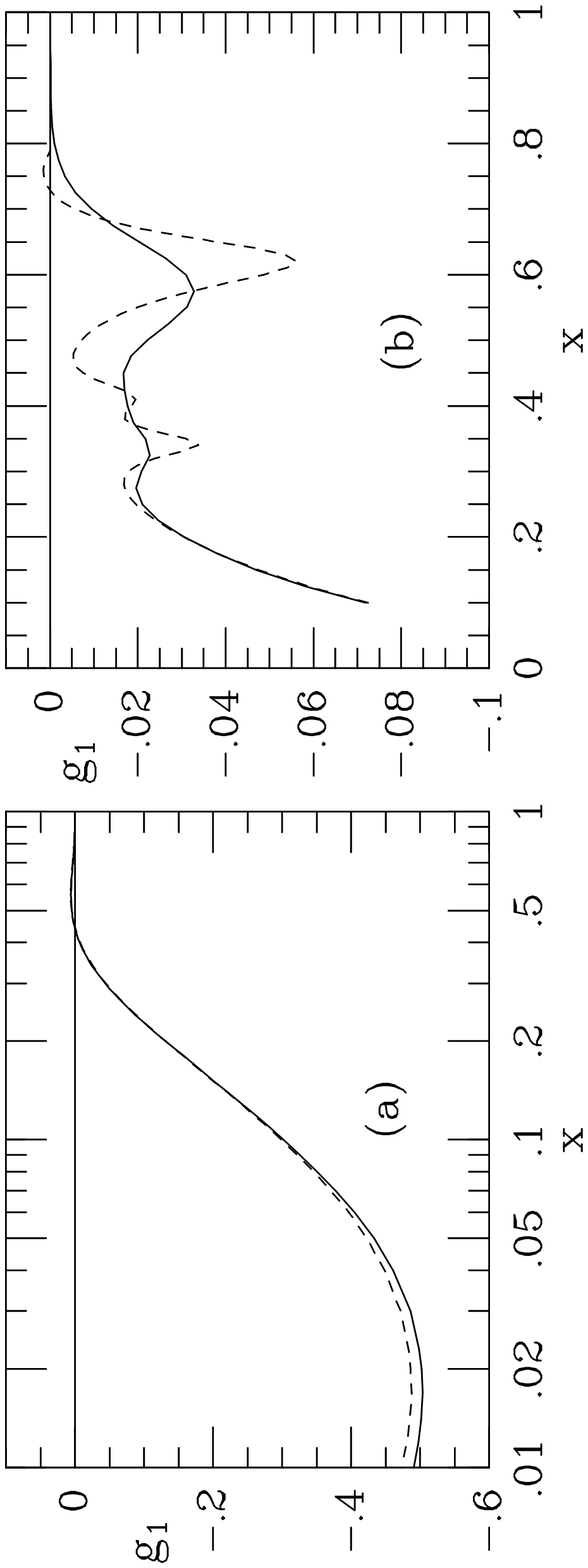}
\end{figure}
\centerline{\large C. Ciofi degli Atti, Phys. Lett. B}
\centerline{\bf FIGURE 1}

\newpage

\begin{figure}[h]
\vspace{12cm}
\includegraphics{ts954.ps}
\end{figure}
\centerline{\large C. Ciofi degli Atti, Phys. Lett. B}
\centerline{\bf FIGURE 2}

\newpage

\begin{figure}[h]
\vspace{12cm}
\includegraphics{incr.ps}
\label{okint}
\end{figure}
\vspace{2cm}
\centerline{\large C. Ciofi degli Atti, Phys. Lett. B}
\centerline{\bf FIGURE 3}


\begin{thebibliography}{9}
\bibitem{drech} D. Drechsel, {\it Prog. Part. Nucl. Phys.} 
34 (1995) 181.
\bibitem{cebp} 
V. Burkert $et$ $al.$, ``Measurement of Polarized Structure
Functions in Inelastic Electron Proton Scattering using the
CEBAF Large Acceptance Spectrometer",
CEBAF Proposal No. 91 -- 023.
\bibitem{cebd} S. E. Kuhn $et$ $al.$, ``The Polarized Structure Function
$g_1^n$ and the $Q^2$ dependence of the Gerasimov -- Drell -- Hearn Sum 
Rule for the Neutron",
CEBAF proposal No. 93 -- 009.
\bibitem{ceb3} 
Z.E. Meziani, ``Measurement of the Neutron ($^3He$) Spin Structure 
Function at Low $Q^2$; a Connection between the Bjorken and the
DHG Sum Rules", 
CEBAF Proposal No. 94 -- 010.
\bibitem{IOFFE} B.L. Ioffe, V.A. Khoze and L.N. Lipatov, Hard
Processes, North-Holland, New York (1983).
\bibitem{DHG66} S.D. Drell and A.C. Hearn, Phys. Rev. Lett. 162
(1966) 1520;\newline S.B. Gerasimov, Yad. Fiz. 2 (1965) 839 [Sov. J.
Nucl. Phys. 2 (1966) 589].
\bibitem{EMC} J. Ashman {\it et al}, Phys. Lett. B 206 (1988) 364;
Nucl. Phys. B 328 (1990) 1.
\bibitem{ik73} I. Karliner, Phys. Rev. D 7 (1973) 2712; R. Workman and
R. Arndt, Phys. Rev. D 45 (1992) 1789; D. Drechsel and M.M. Giannini, 
Few--Body Systems 15 (1993) 99.
\bibitem{burk} V. Burkert and Zhujun Li, Phys. Rev. D 47 (1993) 46;
V. Burkert and B. L. Ioffe, Phys. Lett. B 296 (1992) 223.
\bibitem{ar} MAMI Proposal A2/2 -- 93. Experimental check of the
GDH sum rule, Mainz (1993).
\bibitem{plb} C. Ciofi degli Atti, L.P. Kaptari, S. Scopetta and
A.Yu. Umnikov, Physics Letters B (1996) (to appear).
\bibitem{pg} C. Ciofi degli Atti, S. Scopetta, E. Pace, and G. Salm\`e,
Phys. Rev. C 48 (1993) R968.
\bibitem{sa} R.W. Schultze and P.U. Sauer, Phys. Rev. C 48 (1993) 38.
\bibitem{prc} C. Ciofi degli Atti, E. Pace, and G. Salm\`e,
Phys. Rev. C 46 (1992) R1591.
\bibitem{trial} L.P. Kaptari, A.Yu. Umnikov Phys. Lett. B 240 (1990) 203.
\bibitem{fs} L. Frankfurt, V. Guzey, M. Strikman, 
hep -- ph/9602301 (1996).
\bibitem{wal} R. L. Walker, Phys. Rev. 182 (1969) 1729.
\bibitem{cdayl} C. Ciofi degli Atti, D. Day, and S. Liuti,
Phys. Rev. C 46 (1992) 1045.
\bibitem{kanna}  L.P.  Kaptari, F. Khanna and 
A.Yu.  Umnikov Z.  Phys. A 348 (1994) 211.




\end{thebibliography}
\end{document}